\documentclass[fleqn,12pt,twoside]{article}
 \usepackage{espcrc1}

\readRCS $Id: espcrc1.tex,v 1.2 2004/02/24 11:22:11 spepping Exp $
\usepackage{graphicx}
\usepackage[figuresright]{rotating}

\newcommand{\AmS}{{\protect\the\textfont2
  A\kern-.1667em\lower.5ex\hbox{M}\kern-.125emS}}

\title{Electromagnetic fields and transport coefficients in a hot pion gas}

\author{A.G\'omez Nicola\address[MCSD]{Departamentos
 de F\'{\i}sica Te\'orica I, II,
        Universidad Complutense, 28040 Madrid, Spain}
        \thanks{Work  supported by the Spanish research projects and
        grants FPA2004-02602,FPA2005-02327,
       PR27/05-13955-BSCH, BES-2005-6726.},
        D.Fern\'andez-Fraile\addressmark}

\begin{document}

\maketitle
\begin{abstract}
We present recent results on finite temperature electromagnetic
form factors and the electrical conductivity in a pion gas. The
standard Chiral Perturbation Theory power counting needs to be
modified for transport coefficients. We pay special attention to
 unitarity and to possible applications  for
dilepton and photon production.
\end{abstract}
\section{Introduction}
The most relevant properties of electromagnetic interactions in a
hot and dense medium can be inferred from the retarded
current-current correlator $\Pi^R_{\mu\nu}(\omega,\vec{q})$. On the
one hand, Linear Response Theory provides the reaction of the system
to soft external fields \cite{lebellac}. Thus, the electrical
conductivity measures the response to a constant electric field,
whereas the Debye mass parametrizes the  screening of a single
charge placed at the origin:
\begin{equation}
\sigma (T)=\lim_{\omega\rightarrow
0^+}\lim_{\vert\vec{q}\vert\rightarrow 0^+} \frac{\mbox{Im}
(\Pi^R)^i_i (\omega,\vert\vec{q}\vert)}{3\omega}\quad ; \quad
m_D^2=-\lim_{\vert\vec{q}\vert\rightarrow 0^+}
\lim_{\omega\rightarrow 0^+} \Pi^R_{00}(\omega,\vert\vec{q}\vert)
\label{debsigdef}
\end{equation}

On the other hand, $\mbox{Im}
(\Pi^R)^\mu_\mu(\omega,\vert\vec{q}\vert)$ is directly related to
the photon yield emanated from a Relativistic Heavy Ion Collision
for $q^2=0$ and to the dilepton rate for $q^2=M^2$ with $M$ the
dilepton invariant mass \cite{alam01}.

Here we are interested in a pion gas at finite temperature and zero
baryon density. Below the chiral phase transition, the dynamics of
such a system can be described  with Chiral Perturbation Theory
(ChPT) \cite{gale87}, the most general low-energy expansion
compatible with the spontaneous breakdown of chiral symmetry. The
lagrangian is written as an expansion in pion field derivatives and
masses and Weinberg's chiral power counting \cite{don}  establishes
that the perturbative contribution of a given diagram is of order
$(E/\Lambda_\chi)^D$ (${\cal O}(E^D)$ for short) with
$D=2(N_L+1)+\sum_d (d-2) N_d$, $N_L$ the number of loops, $N_d$ the
number of vertices coming from the $d$-derivatives lagrangian, $E$ a
typical pion energy  and $\Lambda_\chi\simeq$ 1 GeV. Diagrams with
photon lines can be included by counting $e\Lambda_\chi={\cal O}(E)$
and temperature corrections are perturbative for $T$ well below
$T_c\simeq$ 180-200 MeV.
\section{Pion EM form factors and charge distribution}
Only with pion degrees of freedom, the imaginary part of the
current-current correlator entering the dilepton rate is directly
related, to lowest order in $e$, to the modulus of the pion
electromagnetic form factor squared. Physically, this gives the
contribution of the annihilation of two charged pions to form a
dilepton pair, which is dominant for the low energy part of the
spectrum.

The EM form factor at finite temperature has been calculated in
\cite{glp05} in ChPT to one loop. At zero energy, it provides the
spatial Fourier transform of the pion charge distribution
$F(\vert\vec{q}\vert^2)=Q_T(1-\langle r^2\rangle_T
\vert\vec{q}\vert^2/6+\dots)$, where we get for the net pion charge:
\begin{equation}
Q_T=1-\frac{1}{2\pi^2 f_\pi^2}\int_{m_\pi}^\infty dE
\frac{2E^2-m_\pi^2}{\sqrt{E^2-m_\pi^2}}
n_B(E;T)=1-\frac{m_D^2(T)}{2e^2f_\pi^2}
\end{equation}
where the pion decay constant $f_\pi\simeq$ 93 MeV,
$n_B(E;T)=\left[\exp(E/T)-1\right]^{-1}$ and $m_D^2 (T)$ is
calculated to the same order \cite{kap92}. Therefore, the pion
charge is screened in the thermal bath proportionally to the Debye
mass. As a consequence, the pion charge radius $\langle
r^2\rangle_T$ is notably increased from $T>$ 100 MeV \cite{glp05}.
Estimating the deconfinement temperature as $(4\pi/3)\langle
r^2\rangle^{3/2}_{T_d}n_\pi (T_d)=1$ with $n_\pi$ the pion
density, gives $T_{d}\simeq$ 200 MeV, about 65 MeV lower than
using the same estimate with $\langle r^2\rangle_{T=0}$
\cite{kapusta}.

In order to account for the $\rho$-resonance contribution, we have
used  unitarized thermal form factors $F$ and partial waves
$t_{IJ}$ \cite{glp05,dglp02} satisfying  unitarity  in the center
of mass frame:
\begin{equation}
\mbox{Im} F(E;T)=\sigma_T t_{11}(E;T)F^* (E;T) \quad ; \quad
\mbox{Im} t_{11} (E;T)=\sigma_T \vert t_{11} (E;T)\vert^2
\end{equation}
where $\sigma_T=\sqrt{1-4m_\pi^2/E^2}\left[1+2n_B(E/2;T)\right]$ is
the two-pion thermal space factor. The behaviour of $\vert
F\vert^2(E;T)$ shows a clear thermal $\rho$ broadening while
reducing only slightly its mass. This is compatible with recent
dilepton data in the CERN-SPS NA60 experiment, which rules out a
Brown-Rho like dropping mass scenario with almost no broadening
\cite{vhrapp06}. Our results are also consistent with approaches
based on Vector Meson Dominance \cite{rawa00}.
\section{Transport coefficients in ChPT: the
electrical conductivity}
\begin{figure}
\includegraphics[scale=0.8]{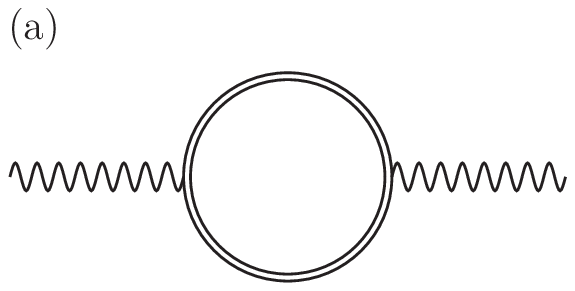}
\hspace*{1cm}
\includegraphics[height=2.6cm]{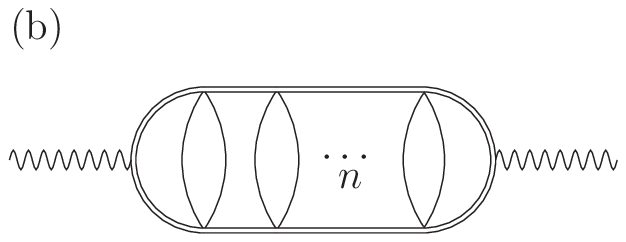}
\vspace*{-0.6cm} \caption{Dominant diagrams for transport
coefficients (a) to one loop, (b) of ladder type.}
\label{diagrams}
\end{figure}

Transport coefficients are intrinsically nonperturbative.
Diagrammatically,  $\mbox{Im} \Pi^R( 0^+,\vec{0})$ shows to one
loop (Figure \ref{diagrams}a)  a typical ``pinching pole"
behaviour $G^R(\omega',\vert \vec{p}\vert)G^A(\omega',\vert
\vec{p}\vert)\approx
\pi\delta\left(\omega'^2-E_p^2\right)/(2E_p\Gamma_p)$, where
$(\omega',\vec{p})$ is the four-momentum carried by the particle
in the loop, $E_p^2=\vert \vec{p} \vert^2+m^2$ and $G^{R,A}$, are
retarded/advanced propagators with the perturbatively small
thermal width $\Gamma_p<<E_p$ \cite{jeon95,valle02}. A dominant
contribution proportional to the inverse width is also expected
from elementary kinetic theory, since transport coefficients are
proportional to mean free times. For instance, the electrical
conductivity $\sigma\sim e^2 N_{ch}/(m\Gamma)$ where $N_{ch}$ is
the number of charge carriers of mass $m$ and width $\Gamma$.

In ChPT, the pion width is  $\Gamma_p={\cal O} (E^5)$\cite{gole89},
so that the above considerations lead to a redefinition of the
standard chiral power counting for the calculation of transport
coefficients \cite{fergn06}. For that purpose, we use the
double-line notation for those internal lines which share the same
four-momentum when the external momentum goes to zero. Those lines
carry propagators with $\Gamma_p\neq 0$ and give the dominant
 contributions mentioned above. Therefore, double lines
do not count as a chiral loop suppression in this new chiral power
counting. Instead, they are assigned a nonperturbative factor $Y$,
which we estimate by calculating the leading order diagram in
Figure \ref{diagrams}a. Thus, we write the contribution of that
diagram to the electrical conductivity  as $\sigma^{(0)}=e^2 m_\pi
Y$. In the standard chiral power counting we would have $Y={\cal
O} (E^2)$. The next step is to identify the dominant diagrams with
this new counting. As it happens in simple scalar theories
\cite{jeon95}, those are ladder diagrams of the type showed in
Figure \ref{diagrams}b, which in our case are ${\cal
O}(E^{2n}Y^{n+1})$ with $n$ the number of rungs. Note that the
loop rungs carry single lines.
\begin{figure}
\begin{center}
\includegraphics*[scale=0.7]{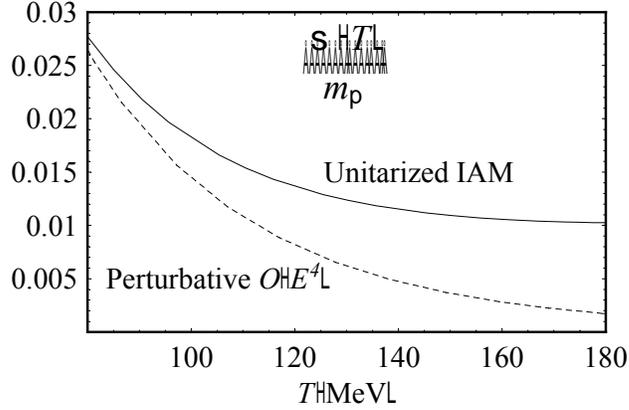}\end{center}
\vspace*{-0.8cm} \caption{The electrical conductivity to leading
order in ChPT, with and without unitarizing the partial waves in
the thermal pion width.} \label{elcond}
\end{figure}

In order to understand the behaviour of the relevant diagrams in
ChPT, let us consider  two different regimes: first, for
$T<<m_\pi$ the typical spatial loop momenta
 $p={\cal O}(\sqrt{m_\pi T})$ and $E\simeq
m_\pi$, we find $Y\simeq \sqrt{m_\pi/T}$, so that  ladder diagrams
could be increasingly important. However,  to the leading
$1/\Gamma$ order, only the spectral functions of the rung loops
contribute \cite{valle02} and every one of those is proportional
to $p_{CM}/m_\pi\sim\sqrt{T/m_\pi}=1/Y$ where $p_{CM}$ is the
center of mass momentum for $\pi\pi$ scattering of the double-pion
(on-shell) lines attached to the rung loop. Therefore,
$\sigma\propto\sqrt{m_\pi/T}$ for very low $T$ and ladder diagrams
give only perturbative corrections to the proportionality
constant. The second regime of interest is $T\sim m_\pi$, where
$p={\cal O}(T)$ and $E\propto T$. Here, unitarization effects have
to be taken into account in the partial waves entering the pion
width. Qualitatively, unitarity makes the conductivity change its
decreasing behaviour with $T$, as showed in Figure \ref{elcond}.
Note that this
 change of behaviour occurs  near $T_c$, which is reasonable since $\sigma$ grows
 with $T$ in the QGP phase \cite{amy00}. A similar behaviour is
expected for other transport coefficients \cite{kapustathis}. In
addition, although ladder diagrams still remain formally
perturbative in ChPT, the presence of derivative vertices
increasing with $T$ makes them numerically important near $T_c$,
where  the corresponding Boltzmann-like integral equations for the
effective vertices \cite{valle02}  have to be solved and so we
will do elsewhere.

Finally, let us comment on a phenomenological application of our
results regarding the photon spectrum. From the definition of the
conductivity in (\ref{debsigdef}), the equilibrium photon rate  at
vanishing energy is proportional to  $T\sigma(T)$ \cite{fergn06}.
This translates directly into a prediction for the photon yield at
 vanishing transverse momentum, where the hadron gas dominates over the QGP.
 Using a simple cylindrical Bjorken expansion with
parameters typical of the CERN WA98 experiment \cite{wa98} gives
$\omega dN_\gamma/d^3\vec{q} (q_T\rightarrow 0^+)\simeq 5.6\times
10^2$ GeV$^{-2}$. This value is reasonably close to a linear
extrapolation of the two  closest  experimental points to the
origin in \cite{wa98} and is also compatible with recent
theoretical analysis \cite{turaga04}. Our result  highlights the
importance of considering a nonzero pion width and  resonances in
$\pi\pi$ scattering for the photon spectrum near zero energy.
\section{Conclusions}
We have presented a recent analysis of electromagnetic properties
of a pion gas. While the pion electromagnetic form factor can be
computed and unitarized in standard ChPT, the calculation of
transport coefficients requires a redefinition of the standard
chiral power counting in order to account for typical
contributions proportional to the inverse particle width.  Our
results show phenomenological and theoretical consistency in the
context of dilepton and photon production at low energies and open
up the possibility of studying in ChPT other physically relevant
transport coefficients such as viscosities.

\end{document}